\documentclass[pra,twocolumn,amsmath,amssymb,superscriptaddress,longbibliography]{revtex4-2}

\usepackage{color,times}
\usepackage{bm}
\usepackage{dcolumn}
\usepackage{multirow}
\usepackage{amssymb,amsfonts,amsmath,graphicx}
\usepackage{epsfig}
\usepackage{xcolor}
\usepackage{braket}
\usepackage{epstopdf}
\usepackage{physics}
\usepackage{stmaryrd}
\usepackage{bbold}
\usepackage{amsmath}
\usepackage[utf8]{inputenc}
\usepackage{chngcntr}
\usepackage[english]{babel}
\usepackage[colorlinks=true, breaklinks=true, linkcolor=darkblue, citecolor=darkblue, urlcolor=darkblue]{hyperref}
\usepackage{textcomp}
\usepackage{soul}
\usepackage{makecell}
\usepackage{todonotes}
\usepackage{siunitx}
\usepackage{bbm}

\definecolor{darkblue}{rgb}{0, 0, 0.8}
\definecolor{darkgreen}{rgb}{0, 0.5, 0}

\newcommand{\quota}[1]{``#1''} % to put quotations use \quota{the text}, it is like writing ``the text''

\begin{document}
	\title{Probing spin-motion coupling of two Rydberg atoms by a Stern-Gerlach-like experiment}

	\author{Gabriel~Emperauger}
	\thanks{GE, MQ, and GB contributed equally to this work.}
	\affiliation{Université Paris-Saclay, Institut d'Optique Graduate School, CNRS, Laboratoire Charles Fabry, 91127 Palaiseau Cedex, France}
	
	\author{Mu~Qiao}
	\thanks{GE, MQ, and GB contributed equally to this work.}
	\affiliation{Université Paris-Saclay, Institut d'Optique Graduate School, CNRS, Laboratoire Charles Fabry, 91127 Palaiseau Cedex, France}
	
	\author{Guillaume~Bornet}
	\thanks{GE, MQ, and GB contributed equally to this work.}
	\affiliation{Université Paris-Saclay, Institut d'Optique Graduate School, CNRS, Laboratoire Charles Fabry, 91127 Palaiseau Cedex, France}
	\affiliation{Department of Electrical and Computer Engineering, Princeton University, Princeton, NJ 08544, USA}
	
	\author{Yuki~Torii~Chew}
	\affiliation{Université Paris-Saclay, Institut d'Optique Graduate School, CNRS, Laboratoire Charles Fabry, 91127 Palaiseau Cedex, France}
	\affiliation{Institute for Molecular Science, National Institutes of Natural Sciences, Okazaki 444-8585, Japan}
	
	\author{Romain~Martin}
	\affiliation{Université Paris-Saclay, Institut d'Optique Graduate School, CNRS, Laboratoire Charles Fabry, 91127 Palaiseau Cedex, France}
	
	\author{Bastien~G\'ely}
	\affiliation{Université Paris-Saclay, Institut d'Optique Graduate School, CNRS, Laboratoire Charles Fabry, 91127 Palaiseau Cedex, France}
	
	\author{Lukas~Klein}
	\affiliation{Université Paris-Saclay, Institut d'Optique Graduate School, CNRS, Laboratoire Charles Fabry, 91127 Palaiseau Cedex, France}
	
	\author{Daniel~Barredo}
	\affiliation{Université Paris-Saclay, Institut d'Optique Graduate School, CNRS, Laboratoire Charles Fabry, 91127 Palaiseau Cedex, France}
	\affiliation{Nanomaterials and Nanotechnology Research Center (CINN-CSIC), Universidad de Oviedo (UO), Principado de Asturias, 33940 El Entrego, Spain}
	
	\author{Thierry~Lahaye}
	\affiliation{Université Paris-Saclay, Institut d'Optique Graduate School, CNRS, Laboratoire Charles Fabry, 91127 Palaiseau Cedex, France}
	
	\author{Antoine~Browaeys}
	\affiliation{Université Paris-Saclay, Institut d'Optique Graduate School, CNRS, Laboratoire Charles Fabry, 91127 Palaiseau Cedex, France}

\date{\today}
	
\begin{abstract}
We propose and implement a protocol to measure the state-dependent motion of Rydberg atoms induced by dipole-dipole interactions. Our setup enables simultaneous readout of both the atomic internal state and position on a one-dimensional array of optical tweezers.
We benchmark the protocol using two atoms in the same Rydberg state, which experience van der Waals repulsion, and measure velocities in agreement with theoretical predictions. When preparing the atoms in a different pair state, we observe an oscillatory dynamics that we attribute to the 
proximity of a macrodimer bound state.
Finally, we perform a Stern-Gerlach-like experiment in which a superposition of the two previous pair states results in the separation of the atomic wavepacket into two macroscopically distinct trajectories, thereby demonstrating spin-motion coupling mediated by the interactions.
\end{abstract}
	
\maketitle
	
\section{Introduction}
The coupling between internal states (\quota{spin} states) and motional degrees of freedom in atomic systems is a resource for many applications in quantum information processing and quantum simulation, such as entangling gates for trapped ions~\cite{CZ_gate_1995, MS_gate_1999}, quantum many-body physics~\cite{Monroe_rmp_2021}, phononic networks~\cite{Chen_nphys_2023}, and quantum state measurements~\cite{Wu_2019}. The textbook situation of spin-motion coupling is the Stern-Gerlach experiment~\cite{Gerlach_Stern_1922}, where spin-dependent forces drive atoms with different internal states along distinct spatial trajectories. 
While such spin-dependent forces have been demonstrated with trapped ions~\cite{Haljan_2005,Monroe_rmp_2021}, Bose-Einstein condensates~\cite{Cronin2009}, and optical lattices~\cite{Salomon2019} by applying external laser or magnetic field gradients, the role of the atomic interactions in generating spin-dependent forces is much less explored experimentally. 
Here, we focus on the state-dependent forces generated between two Rydberg atoms~\cite{Ates_2008,Wuster_2010}.
	
The influence of dipole-dipole interactions between Rydberg atoms on their internal degrees of freedom has been extensively studied~\cite{Reinhard_2008,Beguin_2013,Ravets_2014,Browaeys_2016,Anand_2024,Emperauger_2025} and is now used as a powerful tool for quantum simulations~\cite{Browaeys2020_NP,Ebadi2021,Scholl2021,Chen2023} and quantum computation~\cite{Zhang2020,Bluvstein2024}. 
Their effects on external degrees of freedom have also been measured, for example in Rydberg clouds which expand due to the van der Waals 
repulsion~\cite{Amthor_2007,Teixeira_2015,Thaicharoen_2015,Faoro_2016}.
Signatures of combined effects on internal and external degrees of freedom have been observed for trapped atoms, resulting in the entanglement of the quantized vibrational levels 
(phonons) with the spin degrees of freedom~\cite{Bharti2024,Shaw_2025} or the damping of 
a spin-exchange oscillation~\cite{Chew_2022}. However, a direct measurement of spin-motion coupling for individually-resolved atoms in the absence of external force is lacking: 
in this context, one expects state-dependent dipole-dipole forces to act on the average position, leading to state-dependent trajectories that macroscopically differ.
	
Spin-motion coupling between Rydberg atoms can be a resource for producing entangled 
states~\cite{Mazza2020,Mehaignerie2023,Magoni2023,Bohnmann2024,Wuster_2018,Nill_2025,Parvej_2025}, 
but also as a limiting factor for the fidelity and coherence of quantum information 
processing~\cite{Chew_2022,Mehaignerie2023}. 
In particular, unwanted atomic motion driven by the van der Waals forces can lead to dephasing and heating, presenting challenges for high-precision quantum simulations and scalable quantum computation~\cite{Roghani_2011,Keating_2015,Robicheaux_2021,Zhang_2024,Wyenberg_2025}.
A better understanding of spin-motion coupling is therefore essential for mitigating these effects in the near future and enhancing the performance of Rydberg platforms.

In this work, we report experiments tracking the motion of two Rydberg atoms 
induced by their interactions. We first apply it to the simple case of two atoms prepared 
in the same Rydberg state~$\ket{nS}$, and monitor their separation due to the repulsive van der Waals interaction. 
Second, we prepare a Bell state %$\left(\ket{nS,nP}+\ket{nP,nS}\right)/\sqrt{2}$ 
which experiences an attractive force at large distance, and observe oscillations of the interatomic 
distance which we attribute to the presence of a nearby Rydberg macrodimer. 
Finally, we perform a Stern-Gerlach-like experiment by preparing a superposition of the two previous states, 
leading to the splitting of the atomic positions into two macroscopically distinct trajectories.

%%%%%%%%%%% Figure 1 %%%%%%%%%%%%%%%%
\begin{figure}%[!htb]%\mbox{}
\includegraphics%[width=\linewidth]
{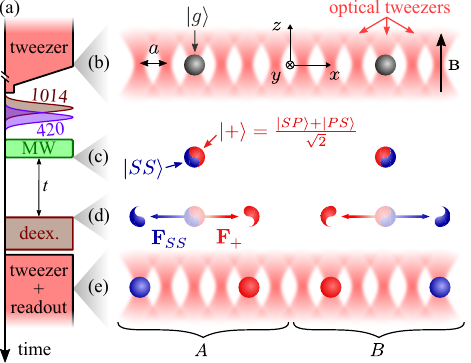}
	\caption{\label{fig:setup} {\bf Experimental sequence}.
	The sequence (a) consists of four steps which are illustrated on the right. 
	(b)~\textit{Loading}: a 1D array of optical tweezers separated by $a \simeq 2$~µm is filled with two ground-state $^{87}\text{Rb}$ atoms (gray balls). The tweezers depth is then adiabatically ramped down by a factor $\sim 100$. 
	(c)~\textit{State preparation}: after switching off the tweezers, a set of optical beams (at~$420$ and $1014$~nm) and microwave pulses (MW at $\sim 17$~GHz) prepare the atoms in the pair states $\ket{S,S}$ (blue balls), $\ket{+} = \left( \ket{S,P} + \ket{P,S} \right)/\sqrt{2}$ (red balls) or in a superposition of both (see text). 
	(d)~\textit{Spin-motion coupling}:  the atoms interact under the Rydberg-Rydberg interactions, without any external force. The forces~$\mathbf{F}_{SS}$ and $\mathbf{F}_{+}$ depend on the internal state, leading to a coupling between internal and external degrees of freedom. 
	(e)~\textit{Readout}: each atom's internal state is measured by imaging the state $\ket{S}$, as well as its position in the optical tweezer. 
	$A$ (resp.~$B$) are the sites on the left (resp. right) part of the array.}
\end{figure}
%%%%%%%%%%%%%%%%%%%%%%%%%%%%%%%%
	
\section{Experimental setup} \label{Sec:setup}
Our experimental sequence, illustrated in Fig.~\ref{fig:setup}, begins by loading two $^{87}\text{Rb}$ atoms in a one-dimensional holographic optical tweezer array generated by a spatial light modulator~\cite{Barredo_2016,Schymik_2020} with $a \approx 2$~µm lattice spacing. The initial distance~$r_0$ between the atoms is thus a multiple of~$a$. In practice, we image the tweezers' light on a diagnostics camera to estimate 
the exact tweezers' positions, which are subject to small random static fluctuations due to optical aberrations (see App.~\ref{SM:static_positional_disorder}). The trapping frequencies of the tweezers are approximately 80~kHz along the transverse directions ($x$ and $y$) and 12~kHz along the axial direction ($z$), with a trapping depth of about 20~MHz. 
The atoms' motional states are initialized through Sisyphus cooling, while their internal states are optically pumped to $\ket{g} = \ket{5S_{1/2},F=2,m_F=2}$.
To set the quantization axis, we then switch on a magnetic field $\mathbf{B}$ along the transverse direction of the tweezers with a magnitude $|\mathbf{B}|=45$~G. We adiabatically reduce the tweezer 
depth by a factor~$\sim 100$, to lower the trapping frequencies by a factor~$10$ and thus reduce the initial velocity dispersion.
Next, we switch off the tweezers and excite the atoms to the Rydberg state 
$\ket{S}=\ket{80S_{1/2},m_J=1/2}$ using a stimulated Raman adiabatic passage 
(STIRAP) through the intermediate state $\ket{i}=\ket{6P_{3/2},F=3,m_F=3}$. 
When required, a microwave (MW) pulse addresses the transition to the state $\ket{P}=\ket{80P_{1/2}, m_J=-1/2}$.
Altogether, Rydberg excitation and microwave pulses typically last a few microseconds, which is much faster than atomic motion.
	
Next, we let the Rydberg atoms interact for a duration~$t$.
During this time, the internal state of the two atoms is described under the Born-Oppenheimer approximation~\cite{Weissbluth_1978,Schwabl_2007} by a basis of electronic eigenstates $\{\ket{\psi_\alpha(\mathbf{r})}\}_\alpha$ parameterized by the internuclear 
distance~$\mathbf{r}$ and quantum numbers~$\alpha$. 
The associated eigenenergies, the Born-Oppenheimer potentials $V_{\alpha}(\mathbf{r})$, are obtained by diagonalizing the two-atom dipole-dipole interaction Hamiltonian~\cite{Weber_2017}.
For each electronic eigenstate~$\ket{\psi_\alpha}$, 
the nuclei follow independent trajectories governed by the Ehrenfest theorem:
\begin{align}
\mu \dfrac{d^2 \langle \hat{\mathbf{r}} \rangle_\alpha}{dt^2} = 
-\left\langle \boldsymbol{\nabla} V_\alpha (\hat{\mathbf{r}}) \right\rangle_\alpha
\label{eq:ehrenfest}
\end{align}
with $\mu$ the reduced mass of the nuclei, $\hat{\mathbf{r}}$ the distance operator and where $\langle \hdots \rangle_\alpha$ stands for the quantum average over internal and external degrees of freedom in the state associated with $\ket{\psi_\alpha}$. 
Further assuming that $\left\langle \boldsymbol{\nabla} V_\alpha (\hat{\mathbf{r}}) \right\rangle_\alpha \simeq \boldsymbol{\nabla} V_\alpha \left(\langle \hat{\mathbf{r}}\rangle_\alpha\right)$ 
(which is valid as long as the quantum fluctuations of $\hat{r} = |\hat{\mathbf{r}}|$, initially on the order of~$0.1$~µm, 
remains much smaller than its average $\langle \hat{r} \rangle_\alpha$), 
the internuclear motion can be interpreted as a classical trajectory 
$\langle \hat{\mathbf{r}} \rangle_\alpha(t)$ under a force $\mathbf{F}_\alpha = -\boldsymbol{\nabla} V_\alpha$.
	
After a duration~$t$, we read out both the internal state and the position of the atoms along the direction~$x$ of the array. 
To do so, we first de-excite the atoms in~$\ket{S}$ to the ground state~$5S_{1/2}$, by coupling $\ket{S}$ to $\ket{i}$ using a 1014~nm laser and letting them spontaneously decay from $\ket{i}$ to $5S_{1/2}$. We then turn the optical tweezer array back on to recapture the atoms in~$5S_{1/2}$ and expel the ones remaining in the Rydberg states. 
Finally, we image the recaptured atoms.  
The presence of an atom indicates that it was initially in $\ket{S}$, 
and the tweezer in which it is recaptured provides a discrete measurement of its position, with a spatial resolution~$a$. The tweezer array is thus used as a \quota{ruler} to measure the atomic positions, similarly to what is done in quantum gas microscopy (e.g. ~\cite{Cheuk_2015,Parsons_2015,Omran_2015,Verstraten_2025}). 
Our method only allows us to image atoms in $\ket{S}$, resulting in the 
loss of positional information for atoms in $\ket{P}$~\footnote{This problem can in principle be compensated by applying a microwave $\pi$-pulse before the deexcitation, in order to recapture atoms in $\ket{P}$. 
However, we did not do it in this study.}.
This whole sequence is repeated typically 1000 times for a given time~$t$ to accumulate statistics.
	
%%%%%%%%%% Figure 2 %%%%%%%%%%%%%%%%%	
\begin{figure*}%[!htb]
\mbox{}
\includegraphics%[width=0.8\textwidth]
{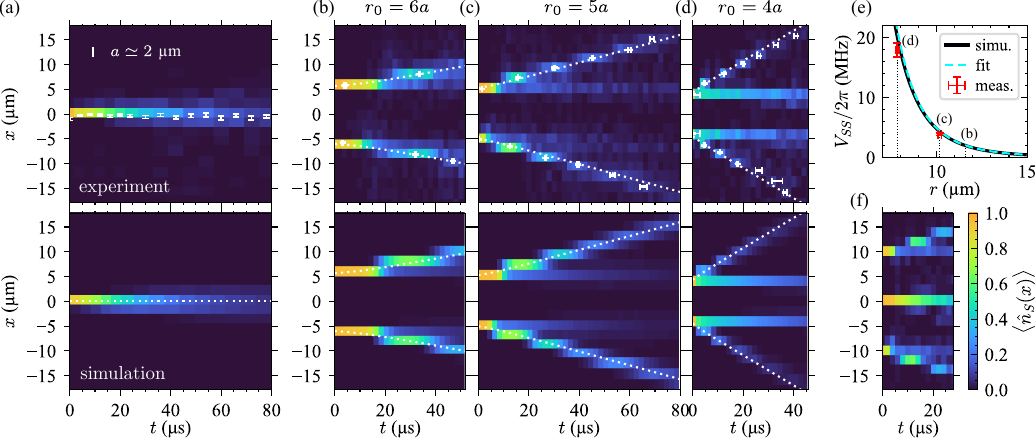}
	\caption{\label{fig:vdw} 
		\textbf{Trajectories of Rydberg atoms under a repulsive van der Waals interaction}, 
		when prepared in the same Rydberg state~$\ket{S}$.
		(a)--(d)~Each panel shows the recapture probability (color scale) per optical tweezer 
		(labeled by their position~$x$ on the vertical axis) as a function of the evolution time~$t$ (horizontal axis). 
		Dotted white lines indicate the predicted 1D classical trajectories under $V_{SS}$. 
		The top panels are experimental data and the bottom panels are numerical simulations 
		taking into account various experimental imperfections (see text). The white data points are the extracted experimental trajectories (see text).
		(a)~Single-atom reference.
		(b,c,d)~Two atoms separated by various initial distances $r_0 \in \{6a, 5a, 4a\}$; 
		the van der Waals repulsion results in symmetric trajectories away from the origin.
		(e)~Calculated potential of the state~$\ket{S,S}$ along~$x$ using~\cite{Weber_2017} (solid black line) 
		and fit by the functional form~$V_{SS}(r) = C_6/r^6$. 
		% giving $C_6 = 2\pi \times (4.85 \pm 0.04)$~THz$\cdot$µm$^6$.
		The red data points indicate the values of $V_{SS}$ extracted from panels~(c) and (d) (see text).
		(f)~Chain of three atoms with nearest-neighbor distance $r_0 = 5a$; the central atom stays nearly fixed due to balanced forces from its neighbors.}
	\end{figure*}
%%%%%%%%%%%%%%%%%%%%%%%%%%%%%
	
\section{Measurement of a repulsive van der Waals force} \label{Sec:repulsive_force}
	
We start by looking at the classical motion of two atoms prepared in the same state $\ket{S}$. 
To rule out possible bias, we first initialize a single atom in $\ket{S}$ 
and monitor its trajectory using the previous protocol. 
Figure~\ref{fig:vdw}(a, top panel) shows the site-resolved recapture probability for atoms in~$\ket{S}$, that is to say the average $\langle \hat{n}_S(x)\rangle(t)$ where $\hat{n}_S(x) = 1$ (resp. 0) indicates the presence (absence) of an atom in~$\ket{S}$ in the optical tweezer at position~$x$. The white data points display the evolution of the center of mass $x_\text{cm}(t) = \int x \cdot \langle \hat{n}_S(x)\rangle(t)dx$, confirming that the atom is on average not deflected. 
However, at times $t>20$~µs, the atom is sometimes recaptured in nearby tweezers, suggesting that the distribution 
of atomic positions is extending. 
To check quantitatively this statement, we perform a numerical simulation of the three-dimensional (3D) 
motion taking into account thermal fluctuations of position, state preparation errors, 
finite Rydberg lifetimes and a finite recapture probability (see App.~\ref{SM:simus_benchmarking}). We find a good agreement with the data using a temperature of $30\pm6$~µK 
(before adiabatic ramp down of the tweezers) as a free parameter [Fig.~\ref{fig:vdw}(a), bottom panel]. 
This temperature means that thermal fluctuations of the distance ($\sim 0.5$~µm at $t=0$) 
dominate over quantum ones ($\sim 0.1$~µm at $t=0$).
	
We now initialize two atoms in $\ket{S}$ at various distances~$r_0 \in \{6a, 5a, 4a\}$ and measure their trajectories using the same sequence [Fig.~\ref{fig:vdw}(b,c,d), top panels]. As time goes by, atoms are predominantly recaptured in optical tweezers further apart, up to six sites away from the initial site, indicating that they move away from each other. 
The smaller the initial distance $r_0$,  the faster the dynamics. 
This behavior originates from the van der Waals potential $V_{SS}(\mathbf{r}) \simeq \hbar C_{6} / r^6$ of the pair state $\ket{S,S}$ with $C_6 = 2\pi \times (4.85 \pm 0.04)$~THz$\cdot$µm$^6$ [Fig.~\ref{fig:vdw}(e)], which results in a repulsive force $\mathbf{F}_{SS}=-\boldsymbol{\nabla} V_{SS} \simeq 6 \hbar C_{6} / r^7 \mathbf{e}_r$ acting on both atoms, $\mathbf{e}_r = \mathbf{r} / r$ being the unit vector along the interatomic axis. 
The ideal classical trajectories, obtained by integrating Eq.~(\ref{eq:ehrenfest}), are plotted in white dotted lines in Fig.~\ref{fig:vdw}(b,c,d) and overlap well with the experimental data. At the smallest initial distance $r_0 = 4a$, each atom experiences an initial force of about~$10^{-20}$~N corresponding to an acceleration of $\sim$70~km$\cdot$s$^{-2}$.

To measure the initial potential energy $V_{SS}(r_0)$, we rely on the conservation of the average energy, which determines the final relative velocity $v_\infty(r_0)$ such that $\mu v_\infty^2/2 = \hbar V_{SS}$ (neglecting the effects of velocity dispersion). 
A similar technique is used in~\cite{Thaicharoen_2015}. 
For each initial distance $r_0 \in \{5a, 4a\}$, we use the following protocol. First, we determine the trajectory of each atom by extracting the time~$t_j$ with the highest recapture probability for each tweezer~$j$ at position $x_j$, resulting in the set of white data points $\{t_j, x_j\}$~\footnote{We could as well have calculated the center of mass of each atom, but the center of mass can be biased by the residual population in the ground state which does not spatially evolve.}. Then, we introduce a cutoff distance $r_c = 9a$ at which $\mathbf{F}_{SS}$ becomes negligible during the time scales at play, and we select the points~$j$ of the trajectory for which $|x_j| \ge r_c / 2$. Finally, we fit the selected data points by a linear function (not shown) to extract $v_\infty(r_0)$, and we plot the resulting values of $V_{SS}(r_0)$ in Fig.~\ref{fig:vdw}(e). We find a good agreement with the theory predictions. For $r_0 = 6a$, the measured trajectory does not contain enough data points to extract a reliable value of $v_\infty$.
	
Finally, in a three-atom configuration [Fig.~\ref{fig:vdw}(f)] with equal spacing~$r_0 = 5a~$ 
between adjacent atoms, the central atom experiences balanced forces from its neighbors and remains static, 
whereas the two atoms on the edges are expelled.
	
Several aspects of the data cannot be explained by the ideal classical trajectory: 
finite spread of the trajectories; 
fading out of the recapture probability at long times; 
residual population at the initial position, particularly visible for $r_0=4a$. 
To understand those deviations, we include experimental imperfections in our two-atom simulations 
(see App.~\ref{SM:simus_benchmarking}) 
using the same parameters as for the single-atom case. 
The results reproduce very well the data [Fig.~\ref{fig:vdw}(b,c,d), bottom panels] 
with the finite excitation probability $1-\eta_\text{STIRAP}$ as the only free parameter 
(see values in Tab.~\ref{tab:values_proba}). 
The latter parameter is needed to explain the residual population at the initial position: 
if one of the two atoms remains in its internal ground state, there is no van der Waals 
repulsion and thus both atoms are recaptured in their initial tweezer. 
This imperfection is more visible as $r_0$ gets smaller, due to the Rydberg blockade of the excitation which leads to a larger value of $\eta_\text{STIRAP}$. %We estimate the blockade radius to be 8~µm.
The measurement fidelity of our protocol is limited by the available trap depth~$U(\mathbf{r}_j)$ 
at the position~$\mathbf{r}_j$ where the atom~$j$ is located when the tweezer is switched back on, 
and by the atomic velocity~$v_j$ at the same time. If the kinetic energy~$K(v_j) = mv_j^2/2$ is 
larger than $U(\mathbf{r}_j)$, the atom escapes from the tweezer and is not recaptured, 
leading to a \quota{false negative} event. 
We estimate that the largest recapture velocity is obtained for 
$K(v_{\text{max}}) = U(\mathbf{r}_j)$, i.e. $v_{\text{max}}(\mathbf{r}_j) = \sqrt{2 U(\mathbf{r}_j) / m}$. 
In the ideal case where the atom is recaptured exactly at the bottom of a tweezer, $U / h \sim 20$~MHz 
and we get $v_{\text{max}} \sim 0.4$~µm/µs, which is larger than the highest measured velocity 
($v\sim 0.3$~µm/µs for $r_0 = 4a$); but in the worst case where $U=0$ 
(in between two tweezers or far away from the chain along $y$ and $z$), the atom is never recaptured. 
Overall, this leads to spatially-dependent losses which, together with the finite Rydberg lifetimes, 
explain why the average recapture probability is not conserved in time.
	
\section{Measurement of an attractive Rydberg-Rydberg force}\label{Sec:attractive_force}
	
\begin{figure}%[!htb]
\mbox{}
\includegraphics[width=\linewidth]{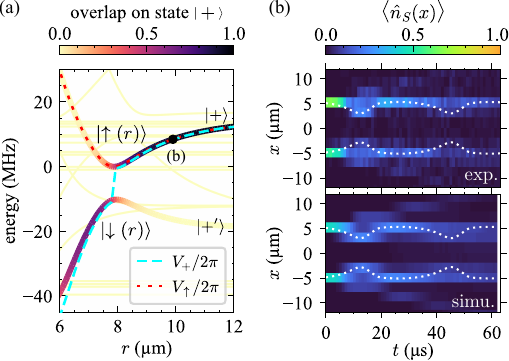}
	\caption{\label{fig:macrodimer_oscillations}
		\textbf{Oscillations around a Rydberg macrodimer.}
		(a)~Calculated Born-Oppenheimer potential curves as a function of interatomic distance, along~$x$. 
		The color bar displays the overlap on the pair state~$\ket{+}$. 
		The dashed blue line is the potential energy~$V_+$ given by effective Hamiltonian theory (see text). 
		The dotted red line~$V_\uparrow$ is the eigenenergy of the state $\ket{\uparrow(r)}$, 
		that governs the adiabatic dynamics of a system initially prepared 
		in $\ket{\uparrow(r)} \simeq \ket{+}$ at $r=5a$ (black dot). 
		It shows an attractive branch that becomes repulsive at short distance, 
		with a minimum corresponding to a Rydberg macrodimer. 
		%The black dot gives the initial condition of the experiment.
		(b)~Trajectory of two atoms prepared in $\ket{+}$: the atoms move inward until $t \sim 12$~µs, 
		then bounce back and partially return to their initial separation by $t \sim 42$~µs. 
		The dotted line is a numerical simulation of the classical motion under $V_\uparrow$.
		The upper (lower) panel in (b) corresponds to the experiment (numerical simulation).}
\end{figure}

To illustrate the dependence of atomic motion on the internal state, 
we now prepare the atoms in the state $\ket{+} = \left( \ket{S,P} + \ket{P,S} \right)/\sqrt{2}$, 
for which the dipolar force is expected to be attractive.
The reason for this is the following. When restricted to the basis 
$\{\ket{S,S}, \ket{S,P}, \allowbreak \ket{P,S}, \ket{P,P}\}$, 
the dipolar interaction simplifies to the effective Hamiltonian
\begin{align}
	H(\mathbf{r}) = \hbar
	\begin{bmatrix}
			V_{S S} & 0 & 0 & 0 \\
			0 & V_{S P} & J_{S P} & 0 \\
			0 & J_{S P} & V_{S P} & 0 \\
			0 & 0 & 0 & V_{P P}
	\end{bmatrix}
	\label{eq:H_eff_two_levels}
\end{align}
where all diagonal terms are van der Waals shifts that scale as $1/r^6$, 
and the off-diagonal term~$J_{SP}$ scales as $1/r^3$. The state $\ket{+}$ is an eigenstate of 
Hamiltonian~(\ref{eq:H_eff_two_levels}) with eigenenergy $\hbar V_+ = \hbar (V_{S P} + J_{S P})$, 
leading to the force $\mathbf{F}_+=-\boldsymbol{\nabla} V_+$. Both $V_{S P}$ and $J_{S P}$ being negative, 
the force is indeed attractive as shown by the blue line in Fig.~\ref{fig:macrodimer_oscillations}(a).

%%%%%%%%%%%% Figure 3 %%%%%%%%%%%	
%%%%%%%%%%%%%%%%%%%%%%%%%%%%
	
To prepare $\ket{+}$, we first initialize two atoms at a distance~$r_0 = 5a$ for which we expect $V_{S P} = -2\pi\times 3.4$~MHz and $J_{S P} = -2\pi \times 3.8$~MHz. 
Then, we excite the pair of atoms to~$\ket{S,S}$ and apply a resonant microwave $\pi$-pulse from $\ket{S,S}$ to~$\ket{+}$, using a Rabi frequency~$\Omega=2\pi\times1.8$~MHz and a detuning $\delta = J_{S P} + V_{S P} - V_{S S} = -2\pi \times 12$~MHz from the single-atom resonance~\cite{de_Leseleuc_2017}.
We use a Gaussian pulse with a $1/e^2$ half-width of 160~ns, much shorter than atomic motion.
The resulting trajectory is shown in Fig.~\ref{fig:macrodimer_oscillations}(b, top panel). The initial value of the recapture probability in the occupied tweezers is now $\langle \hat{n}_S\rangle(t=0) \simeq 0.5$ instead of~$1$, because half of the atomic population is now in~$\ket{P}$.
At a time~$t\sim 12$~µs, both atoms have moved by $\sim 2$~µm towards each other, confirming the attractive nature of the potential. However, at $t \sim 42$~µs the two atoms are back to their initial position. 
To understand this effect, we need to go beyond the effective Hamiltonian picture of Eq.~(\ref{eq:H_eff_two_levels}), which assumes that $\ket{+}$ is well-separated in energy from other pair states. 
Using exact diagonalization of the dipolar interaction~\cite{Weber_2017}, 
we show in Fig.~\ref{fig:macrodimer_oscillations}(a) the Born-Oppenheimer potentials in the vicinity of $\ket{+}$. 
We find that $\ket{+}$ becomes degenerate with the other pair state $\ket{+'} = (\ket{S', P'} + \ket{P', S'}) / \sqrt{2}$, where $\ket{S'} = \ket{80S_{1/2},m_J=-1/2}$ and $\ket{P'} = \ket{80P_{3/2},m_J=-3/2}$. 
Since $\ket{+}$ and $\ket{+'}$ are coupled by the dipole-dipole Hamiltonian, 
this results in new eigenstates $\ket{\uparrow(r)}$ and $\ket{\downarrow(r)}$ 
which are superpositions of $\ket{+}$ and $\ket{+'}$, and whose eigenenergies form an avoided crossing at a distance $r=7.9$~µm.
At this distance, the upper energy branch~$V_\uparrow$ [dotted red line in 
Fig.~\ref{fig:macrodimer_oscillations}(a)] reaches a minimum that corresponds 
to a molecular bound state or so called \quota{Rydberg macrodimer}~\cite{Hollerith_2023}.
	
Assuming that the Born–Oppenheimer approximation holds, 
we simulate the adiabatic motion under $V_\uparrow$ [dotted white line in 
Fig.~\ref{fig:macrodimer_oscillations}(b)]. The resulting oscillation overlaps well with the data. 
In App.~\ref{SM:adiabaticity}, we check that the adiabatic approximation 
is justified by the large energy gap compared with the speed at which the atoms reach the avoided crossing. 
To further confirm this hypothesis, we compare the data with a numerical simulation 
that accounts for experimental imperfections and find an excellent agreement [Fig.~\ref{fig:macrodimer_oscillations}(b, bottom panel)]. 
The residual population that is expelled is due to the imperfect microwave transfer from $\ket{S,S}$ to $\ket{+}$.

%%%%%%%%%%%%%%%%%%% Figure 4 %%%%%%%%%%%%%%%%%%%%	
\begin{figure}%[!htb]
\mbox{}
\includegraphics{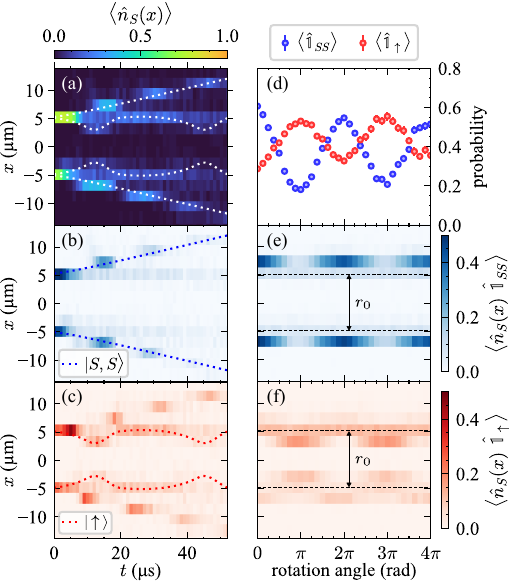}
	\caption{\label{fig:stern-gerlach}
		\textbf{Stern-Gerlach-like experiment}.
		(a,b,c)~Trajectory of two atoms prepared in the  superposition
		$\tfrac{1}{\sqrt{2}}(\ket{S,S}+\ket{+})$, analyzed with: 
		(a)~the site-resolved recapture probability~$\left\langle \hat{n}_S(x) \right\rangle$; 
		(b)~the site-resolved recapture probability correlated with experimental shots 
		where two atoms are recaptured, $\left\langle \hat{n}_S(x) \hat{\mathbbm{1}}_{SS}\right\rangle$; 
		(c)~the site-resolved recapture probability correlated with experimental shots 
		where one out of the two atoms is recaptured, i.e. $\left\langle \hat{n}_S(x) \hat{\mathbbm{1}}_{\uparrow}\right\rangle$.
		The dotted lines are classical simulations of the trajectories.
		(d,e,f)~Rabi oscillation between the states~$\ket{S,S}$ and~$\ket{\uparrow}$ followed by an evolution time~$t=14$~µs, analyzed with: 
		(d)~the total probability of the events~$\hat{\mathbbm{1}}_{SS}$ and~$\hat{\mathbbm{1}}_{\uparrow}$, summed over all tweezers;
		(e)~the site-resolved version $\left\langle \hat{n}_S(x) \hat{\mathbbm{1}}_{SS}\right\rangle$;
		(f)~$\left\langle \hat{n}_S(x) \hat{\mathbbm{1}}_{\uparrow}\right\rangle$.}
\end{figure}
%%%%%%%%%%%%%%%%%%%%%%%%%%%%%%%%%%%%%%%%%
	
\section{A Stern-Gerlach-like experiment} \label{Sec:stern_gerlach}
	
Finally, we realize a Stern-Gerlach-like experiment by applying a $\pi/2$-pulse between $\ket{S,S}$ and $\ket{+}$ instead of a $\pi$-pulse, thus creating the superposition $(\ket{S,S}+\ket{+})/\sqrt{2}$ (Fig.~\ref{fig:setup}). As shown in Fig.~\ref{fig:stern-gerlach}(a), two trajectories emerge from the average recapture probability: the repulsive one corresponding to $\ket{S,S}$ [as in Fig.~\ref{fig:vdw}(c)] and the oscillatory one corresponding to $\ket{\uparrow}$ [as in Fig.~\ref{fig:macrodimer_oscillations}(b)].
In the following, we check the correspondence between those trajectories and the internal states. To do so, we first split the array in two regions~$A$ and $B$, defined in Fig.~\ref{fig:setup}, such that each of those regions contains one of the two atoms. We notice that the internal states $\ket{S,S}$ and $\ket{\uparrow}$ can be distinguished by looking at the number of recaptured atoms: if the system is in $\ket{S,S}$, two atoms should be imaged (one in region~$A$ and one in~$B$); we call the associated event~$\hat{\mathbbm{1}}_{SS}$. Conversely, if the system is in $\ket{\uparrow}$, only one atom should be imaged (either in~$A$ or in~$B$), leading to the event $\hat{\mathbbm{1}}_{\uparrow}$~\footnote{This statement, which is obvious for the state~$\ket{+}$, is also satisfied by $\ket{+'}$ because $\ket{S'}$ is imaged and $\ket{P'}$ is lost during the readout. So it is also true for $\ket{\uparrow}$, which is a superposition of $\ket{+}$ and $\ket{+'}$.}.
Then, we plot in Fig.~\ref{fig:stern-gerlach}(b) the site-resolved recapture probability correlated with the event~$\hat{\mathbbm{1}}_{SS}$, that is to say the quantity~$\left\langle \hat{n}_S(x) \hat{\mathbbm{1}}_{SS}\right\rangle$.
This analysis procedure selects predominantly the repulsive trajectory, as expected when the system is in~$\ket{S,S}$.
%(with a residual immobile background that we again attribute to Rydberg excitation errors). 
Complementary, Fig.~\ref{fig:stern-gerlach}(c) shows the site-resolved recapture probability correlated with the event~$\hat{\mathbbm{1}}_{\uparrow}$. 
This time, the oscillatory behavior appears as predicted; but the repulsive trajectory is also unexpectedly populated. We attribute this effect to atom losses, due to finite Rydberg lifetimes and detection errors: starting from the expelled state~$\ket{S,S}$ (which should lead to the event~$\hat{\mathbbm{1}}_{SS}$ all the time), the loss of an atom will produce an event~$\hat{\mathbbm{1}}_{\uparrow}$ that biases the measurement.

To consolidate our results, we scan the duration of the microwave pulse, thus preparing various superpositions of the states~$\ket{S,S}$ and $\ket{+}$, and we let the system evolve during~$t=14$~µs. This time is chosen to be half the period of the oscillatory trajectory, when losses due to finite Rydberg lifetimes are still small ($\sim5$~\% per atom). Figure~\ref{fig:stern-gerlach}(d) shows the total probabilities of the events~$\hat{\mathbbm{1}}_{SS}$ and $\hat{\mathbbm{1}}_{\uparrow}$, revealing a coherent oscillation between $\ket{S,S}$ and $\ket{\uparrow}$. 
In Fig.~\ref{fig:stern-gerlach}(e,f), we plot the associated site-resolved probabilities and  confirm the expected state-dependent motion, where $\ket{\uparrow}$ has moved inward by approximately one lattice site while $\ket{S,S}$ has propagated outward by one site.

\section{Conclusion}
In summary, we have demonstrated a simple protocol to
track the state-dependent motion of Rydberg atoms induced by their dipole-dipole interactions. 
This approach enables the observation of a wide range of behaviors, from van der Waals repulsion 
to oscillatory dynamics in a macrodimer bound state. It is easily scalable to a larger number of atoms.
The regime of Rydberg spin-motion coupling explored here differs from previous studies 
in that it probes the motion of individual atoms in free space, rather than Rydberg clouds or trapped atoms.

By directly probing how internal spin states translate into atomic motion, our work provides a new way to characterize and quantify spin-motion coupling in Rydberg systems. This opens the door to experimentally realizing many proposals for controlling atomic motion~\cite{Ates_2008}, for studying molecular processes such as non-adiabatic dynamics on conical intersections~\cite{Yarkony_1996,Wuster_2011,Kiffner_2013,Leonhardt_2014,Gambetta_2021}, and for probing many-body phenomena with coupled spin-motion dynamics in Rydberg aggregates~\cite{Schempp_2014,Wuster_2018,Aliyu_2018}. Moreover, our method offers a deeper understanding of these effects, which may ultimately aid in mitigating unwanted motional decoherence in quantum simulation and computation platforms based on Rydberg atoms~\cite{Mitra2020, Mohan2023}.

Our final experiment is a Stern-Gerlach-like scheme mediated  by the interactions, 
leading to the separation of the atomic wavepacket into two macroscopically distinct sets of trajectories. 
In interferometric terms, the microwave pulse combined by the dipole-dipole forces is equivalent to a beam splitter. 
A natural next step would be to investigate the coherence between the attractive 
and repulsive parts of the wavefunction, with the goal of accessing their relative phase or 
exploring entanglement properties. Achieving phase-sensitive measurements would require 
closing the interferometer with a second beam-splitting operation, to recombine the two parts 
of the wavefunction.

\begin{acknowledgments}
This work is supported by the Agence Nationale de la Recherche (ANR-22-PETQ-0004 France 2030, project QuBitAF), 
the European Research Council (Advanced grant No. 101018511-ATARAXIA), 
and the Horizon Europe programme HORIZON-CL4- 2022-QUANTUM-02-SGA (project 101113690 (PASQuanS2.1). 
R.M. acknowledges funding by the “Fondation CFM pour la Recherche” via a Jean-Pierre Aguilar PhD scholarship. 
D. B. acknowledges support from MCIN/AEI/10.13039/501100011033 
(PID2020119667GA-I00, CNS2022-135781, EUR2022-134067 
and European Union NextGenerationEU PRTR-C17.I1).
\end{acknowledgments}

\setcounter{figure}{0}
\renewcommand\thefigure{S\arabic{figure}}
\appendix

\section{Calibration of the tweezers' positions}\label{SM:static_positional_disorder}

%%%%%%%%%% Figure S1 %%%%%%%%%%%%	
\begin{figure}[!b]
	%\mbox{}
	\includegraphics{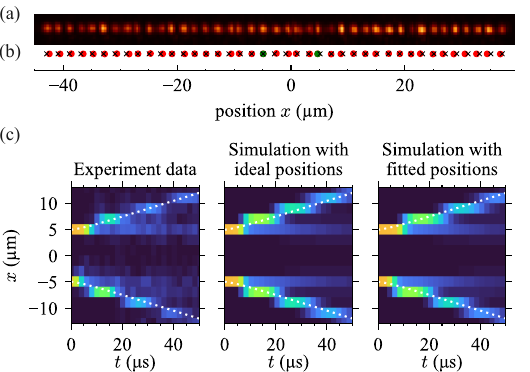}
	\caption{\label{fig:trap_position} 
		\textbf{Estimation of the optical tweezers' positions.}
		(a)~Intensity distribution of the tweezers, imaged on a diagnostics camera after the vacuum chamber.
		(b)~Ideal positions (black crosses) and fitted positions (colored disks). 
		The green disks indicate the initial positions in the benchmark experiment described in Fig.~\ref{fig:vdw}(c), 
		in which two atoms initially separated by $r_0 = 5a$ evolve under a repulsive van der Waals interaction.
		(c)~Benchmark experiment [same data as Fig.~\ref{fig:vdw}(c)] compared to 
		simulations assuming either ideal or disordered tweezers' positions.}
\end{figure}
%%%%%%%%%%%%%%%%%%%%%%%%%%%

The lattice spacing between the tweezers, $a \approx 2$~µm, is comparable to their waist $w = 1.2$~µm (diffraction limit for our numerical aperture $\text{NA} = 0.5$). 
Optical interferences between neighboring tweezers thus lead to random static fluctuations in their positions, on the order of $0.3$~µm~\cite{nishimura2024}. 
In principle, these fluctuations could be eliminated by either using an independent, in-situ measurement to provide feedback to the spatial light modulator for atomic position correction~\cite{Chew_2024,Bornet_PhD_2024}, or by replacing the holographic tweezer array with an optical lattice~\cite{Verstraten_2025}.
Here, we describe the procedure used to estimate the actual positions of the tweezers.

A diagnostics camera located after the vacuum chamber records the intensity profile of the optical tweezers [Fig.~\ref{fig:trap_position}(a)]. We fit each tweezer profile with a 2D Gaussian function to extract its center position, with an average fitting uncertainty of $0.04$~µm.
These fitted positions are shown in Fig.~\ref{fig:trap_position}(b). 
From them, we calculate the standard deviation of the distances between nearest neighbor tweezers to be $0.27$~µm. 
%For all the numerical simulations presented in this work, we use the measured positions and find that they systematically improve the quantitative agreement with the measured recapture probability.
In all numerical simulations presented here, incorporating the experimentally measured atomic positions enhances the agreement with the measured recapture probability, leading to an average $\sim 30\%$ reduction in the reduced $\chi^2$ deviation.

Taking theses static random fluctuations in the tweezers' positions is necessary to reproduce the small deviations  between the measured trajectories and a simulation assuming ideal positions.
In particular, we observe a slight asymmetry between the trajectories of the two atoms. 
As an illustration, Fig.~\ref{fig:trap_position}(c) shows the case of  atoms initially separated by $r_0 = 5a$ 
under a repulsive van der Waals interaction [the same experiment as in Fig.~\ref{fig:vdw}(c)]: there, the maximum recapture probability for the bottom atom, in a tweezer displaced by $-a$ from its initial position, does not occur at the same time as for the top atom, which is recaptured in a tweezer displaced by $+a$. 
Figure~\ref{fig:trap_position}(c) shows two simulations of the trajectories assuming ideal positions of the tweezers or the positions obtained from the fits. The experimental asymmetry is reproduced when including the disorder in the positions: since the distance between the nearest neighbor traps varies, 
the time at which each atom is recaptured is different for two traps that should be symmetrically located.

\section{Numerical simulations with experimental imperfections}\label{SM:simus_benchmarking}

%%%%%%%%%%%%%% Figure S2 %%%%%%%%%%%%%%	
\begin{figure}%[!htb]
	%\mbox{}
	\includegraphics{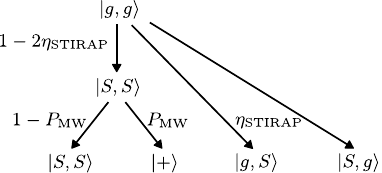}
	\caption{\label{fig:probability_tree} 
		\textbf{Probability tree used to model state preparation errors} in the case of two atoms 
		that should be prepared in $\ket{+}$. 
		The branch with second-order weight in $\eta_\text{STIRAP}$ is neglected.}
\end{figure}
%%%%%%%%%%%%%%%%%%%%%%%%%%%%%%%%%

%%%%%%%%%%%% Table S1 %%%%%%%%%%%%%%%% 	
\begin{table}[]
	\begin{tabular}{c|c|c|c}
		Data set & $\eta_\text{STIRAP}$ & $P_\text{MW}$ & $r_0$ (from camera) \\ \hline \hline
		$\ket{S,S}$ at $r_0 = 6a$ [Fig.~\ref{fig:vdw}(b)] & 0.05$^*$ & 0 & 11.6~µm \\ \hline
		$\ket{S,S}$ at $r_0 = 5a$ [Fig.~\ref{fig:vdw}(c)] & 0.1$^*$ & 0 & 10.2~µm \\ \hline
		$\ket{S,S}$ at $r_0 = 4a$ [Fig.~\ref{fig:vdw}(d)] & 0.3$^*$ & 0 & 7.8~µm \\ \hline
		$\ket{+}$ at $r_0 = 5a$ [Fig.~\ref{fig:macrodimer_oscillations}(b)] & 0.1 & 0.85$^*$ & 10.2~µm \\
	\end{tabular}
	\caption{\label{tab:values_proba} 
		\textbf{Values used in the numerical simulations} for the probabilities defined in 
		Fig.~\ref{fig:probability_tree} and for the exact value of the initial distance~$r_0$, 
		when calibrated from the diagnostics camera (see App.~\ref{SM:static_positional_disorder}). Free parameters are indicated by an asterisk~($^*$).}
\end{table}
%%%%%%%%%%%%%%%%%%%%%%%%%%%%%%%%%

The numerical simulations used for comparison with the data work as follows. 
We consider a system of two atoms, each one with four internal states: 
the two Rydberg states of interest ($\ket{S}=\ket{80S_{1/2},m_J=1/2}$ and $\ket{P}=\ket{80P_{1/2},m_J=-1/2}$), 
the internal ground state $\ket{g}=\ket{5S_{1/2}}$ and another generic Rydberg state $\ket{r}$. 
Although not contributing to the ideal dynamics, the latter two states are required to 
account for state preparation and measurement errors, as well as decoherence. 
The simulation of the one-atom case is a straightforward adaptation 
from the case of two atoms (no interactions).

\noindent
{\bf State initialization.} 
We initialize the system in a statistical mixture of internal pair states $\ket{\psi_\alpha}$ with probabilities taken from 
the tree in Fig.~\ref{fig:probability_tree}. First, each atom has a 
finite probability~$\eta_\text{STIRAP}$ to be left in the ground state instead of being 
excited to $\ket{S}$. Then, the microwave pulse that transfers $\ket{S,S}$ to $\ket{+}$ 
has a finite efficiency $P_\text{MW}$. The values of the probabilities~$\eta_\text{STIRAP}$ 
and $P_\text{MW}$ are given in Tab.~\ref{tab:values_proba}.
	
The initial distance~$\mathbf{r}_0$
is sampled from a thermal distribution at a 
temperature of $30$~µK, corresponding to a standard deviation of $0.48$~µm 
in the transverse direction of the tweezers ($x$ and $y$) and $3.1$~µm in the 
longitudinal direction ($z$). The standard deviation for the initial relative velocity~$\mathbf{v}_0$ 
is $0.024$~m$/$s in all three directions. By modeling the atomic motion as a thermal mixture 
of classical trajectories rather than coherent wavepackets, we neglect the contribution of 
quantum position fluctuations which have a smaller magnitude.

\begin{figure}%[!htb]
	%\mbox{}
	\includegraphics{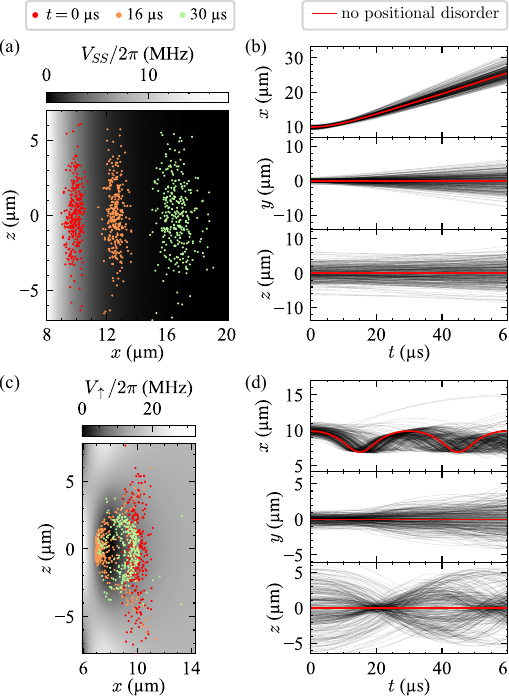}
	\caption{\label{fig:trajectories} 
		\textbf{Simulated trajectories under state-dependent three-dimensional potentials, including positional disorder.}
		(a)~The colormap displays the potential felt by the pair state~$\ket{S,S}$ at position $y=0$. 
		The colored points represent the distribution of atomic distances~$\mathbf{r}(t)$ evolving 
		under~$V_{SS}$ at times $t\in\{0,16,30\}$~µs (projected on the $xz$ plane). 
		The initial average distance is $5a$ along $x$.
		(b)~Time evolution of the atomic distances~$\mathbf{r}(t)$ when projected on the $x$, $y$ and $z$ axes. 
		The red lines highlight the trajectory without positional disorder: 
		$\mathbf{r}_0=5a \mathbf{e}_x$, $\mathbf{v}_0=0$~m$/$s.
		(c,d)~Same as (a,b) in the case of the pair state~$\ket{\uparrow}(\mathbf{r})$, 
		assuming an adiabatic evolution (see App.~\ref{SM:adiabaticity}).}
\end{figure}

\noindent 
{\bf Time evolution.} 
For each initial condition $\{\ket{\psi_\alpha}, \mathbf{r}_0, \mathbf{v}_0\}$, we solve the classical 
equation of motion given by the Ehrenfest theorem (Newton's law)
\begin{align}
	\mu \frac{d^2 \mathbf{r}(t)}{dt^2} = - \boldsymbol{\nabla} V_\alpha\left[\mathbf{r}(t)\right]
\end{align}
where $V_\alpha(\mathbf{r})$ is the state-dependent 3D Born-Oppenheimer potential obtained from~\cite{Weber_2017}. 
Examples of trajectories are shown in Fig.~\ref{fig:trajectories}(a) for $\ket{\psi_\alpha} = \ket{S,S}$ and in 
Fig.~\ref{fig:trajectories}(b) for $\ket{\psi_\alpha} = \ket{\uparrow}$, using 300 samples. 
Then, the position of atom~$1$ (respectively atom~$2$) is chosen to be 
$\mathbf{r}_1 = -\mathbf{r}/2$ (resp. $\mathbf{r}_2 = \mathbf{r}/2$).

%%%%%%%%%%%% Figure S3 %%%%%%%%%%%%%

We include the effect of decoherence induced by the finite Rydberg lifetimes by 
accounting for four single-atom decay channels with associated lifetimes
\begin{align}
	\begin{array}{lll}
		\tau_{\ket{S}\rightarrow\ket{g}} = 643~\text{µs}, & & \tau_{\ket{S}\rightarrow\ket{r}} = 290~\text{µs},\\
		\tau_{\ket{P}\rightarrow\ket{g}} = 1157~\text{µs}, & & \tau_{\ket{P}\rightarrow\ket{r}} = 293~\text{µs}.
	\end{array}
\end{align}
We perform a quantum Monte-Carlo simulation for the time evolution of the internal state~\cite{Dalibard_1992}: 
at every time step $dt \sim 10$~µs, every atom~$j$ has a probability $dP_{\ket{i}\rightarrow\ket{f}}^{(j)}$ 
to decay from state $\ket{i} \in \{\ket{S}, \ket{P}\}$ to state $\ket{f} \in \{\ket{g}, \ket{r}\}$ given by
\begin{align}
	dP_{\ket{i}\rightarrow\ket{f}}^{(j)}(t) = \frac{dt \; P^{\ket{i}}_{j}(t)}{\tau_{\ket{i}\rightarrow\ket{f}}}
\end{align}
with $P^{\ket{i}}_{j}$ the population of atom~$j$ in state~$\ket{i}$. 
If a jump occurs, the dipole-dipole interaction between the two atoms is switched off ($V_\alpha = 0$) 
so that each atom continues its motion in free flight.
	
Finally, the experimental sequence contains a dead time of $4.5$~µs (measured on a photodiode) 
between the deexcitation and the time when the traps are switched on, during which the atoms evolve in free flight. 
To match the data, the simulation also needs to be offset by $1.5$~µs, 
which corresponds to the measured duration between the end of the STIRAP and the deexcitation in the sequence.

\noindent
{\bf Readout.}
Ideally, our protocol would measure all atoms in $\ket{S}$ (and only atoms in this state) at their final position. 
However, several physical effects reduce the measurement fidelity.
	
First, atoms are lost if their kinetic energy exceeds the available trapping potential at their final position. 
To evaluate the recapture probability of an atom~$j$ in an optical tweezer~$k$ at time~$t$, 
we compute the statistical frequency of events where an atom is recaptured:
\begin{align}
	P^\text{recap}_{j,k}(t) = \overline{\mathbbm{1}_{K[\mathbf{v}_j(t)] < U_k[\mathbf{r}_j(t)]} }
\end{align}
where $K(\mathbf{v}) = m \mathbf{v}^2 / 2$ is the single-atom kinetic energy, 
$U_k[\mathbf{r}]$ is the potential of tweezer~$k$ and $\overline{\vphantom{A}\ldots}$ 
stands for the average over positional disorder. We consider Gaussian potentials 
with a waist of $1.2$~µm, that are cut at transverse distances larger than 
half the tweezer spacing ($a/2 \simeq 1$~µm), such that two neighboring tweezers do not overlap.
	
Second, the probability for an atom~$j$ to be imaged depends on the population 
of the internal states (defined before the deexcitation pulse) as
\begin{align}
	P^\text{imaged}_{j}(t) = (1-\varepsilon_S) P^{\ket{S}}_{j}(t) + \varepsilon_P P^{\ket{P}}_{j}(t) + P^{\ket{g}}_{j}(t).
	\label{eq:imaging_proba}
\end{align}
Here, $\varepsilon_S = 5$~\% is the probability that an atom in $\ket{S}$ is not imaged 
(due to residual mechanical losses and the finite fidelity of the deexcitation pulse) 
---false negative events--- and $\varepsilon_P$ is the probability that an atom in $\ket{P} = 4$~\% 
is imaged (due to spontaneous emission before the atom is expelled from the traps) ---false positive events. In the specific case of the data set with initial state~$\ket{+}$ (Fig.~\ref{fig:macrodimer_oscillations} of the main text), where the atoms also have an overlap with the internal states~$\ket{S'}$ and $\ket{P'}$ (see App.~\ref{SM:adiabaticity}), we assume that $\ket{S'}$ and $\ket{P'}$ are detected with the same efficiencies as $\ket{S}$ and $\ket{P}$, allowing us to apply Eq.~(\ref{eq:imaging_proba}) during the whole dynamics.
	
Finally, the total recapture probability in a tweezer~$k$ is
\begin{align}
	P_\text{tot}^{(k)}(t) = \sum_{j \in \{1,2\}} P^\text{recap}_{j,k}(t) \cdot P^\text{imaged}_{j}(t).
\end{align}
where the sum runs over the two atoms~$j$ (but in practice the trajectories of the two atoms never overlap).

%%%%%%%%%%%%%%%%%% Figure S4 %%%%%%%%%%%%%%%%	
\begin{figure}%[!t]
	%\mbox{}
	\includegraphics{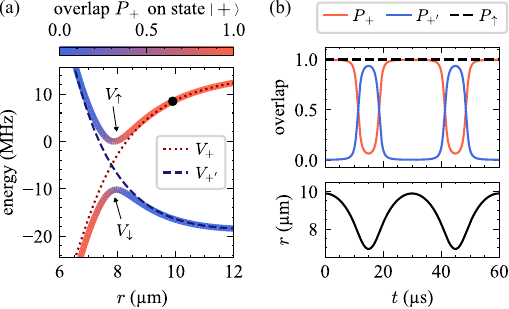}
	\caption{\label{fig:adiabaticity} 
		\textbf{Adiabaticity of the dynamics at the avoided crossing}.
		(a)~State decomposition of the simplified internal Hamiltonian $H_\text{int}$ (see text)
		onto the pair states $\ket{+}$ and $\ket{+'}$, showing an avoided crossing at distance $r=7.9$~µm. 
		The zero of energy is chosen as the minimum of the upper branch $V_\uparrow(r)$. 
		The black dot gives the initial condition of the simulation at $r_0 = 5a$.
		(b)~Time evolution of the internal state (top panel) and of the external state (bottom panel).}
\end{figure}
	
\section{Adiabaticity of the dynamics at the avoided crossing}\label{SM:adiabaticity}
	
In the main text, we assumed that the dynamics of the internal state shown in
Fig.~\ref{fig:macrodimer_oscillations} is adiabatic, meaning that the two-atom system remains 
in the instantaneous eigenstate corresponding to the upper energy branch of the avoided 
crossing~$V_\uparrow$. This is equivalent to the Born-Oppenheimer approximation, 
where the global atomic motion is decoupled from the faster electronic dynamics. 
This section aims at verifying this assumption.

We consider a simplified internal structure composed of the two pair states that are involved 
in the avoided crossing: $\ket{+} = (\ket{S}+\ket{P})/\sqrt{2}$ and $\ket{+'} = (\ket{S'}+\ket{P'})/\sqrt{2}$ with
\begin{align}
	%\left\{
	\begin{array}{l}
		\ket{S}=\ket{80S_{1/2},m_J=1/2} \\
		\ket{P}=\ket{80P_{1/2},m_J=-1/2} \\
		\ket{S'} = \ket{80S_{1/2},m_J=-1/2} \\
		\ket{P'} = \ket{80P_{3/2},m_J=-3/2}.
	\end{array}
	%\right.
	\label{eq:V_up_down}
\end{align}
The internal Hamiltonian in the basis $\{\ket{+},\ket{+'}\}$ is
	\begin{align}
		H_\text{int}(r) = \hbar
		\begin{bmatrix}
			V_{+}(r) & W(r) \\
			W(r) & V_{+'}(r)
		\end{bmatrix}
		\label{eq:H_eff}
	\end{align}
whose eigenstates, $\ket{\uparrow(r)}$ and $\ket{\downarrow(r)}$, have respective eigenenergies (in units of $\hbar$)
	\begin{align}
		%\left\{
		\begin{array}{ll}
			V_\uparrow = \dfrac{V_+ + V_{+'}}{2} + \sqrt{\left(\dfrac{V_+ - V_{+'}}{2}\right)^2 + W^2} \\
			V_\downarrow = \dfrac{V_+ + V_{+'}}{2} - \sqrt{\left(\dfrac{V_+ - V_{+'}}{2}\right)^2 + W^2}.
		\end{array}
		%\right.
		\label{eq:V_up_down}
	\end{align}
To build $H_\text{int}$, we make use of the results from exact diagonalization on a basis of many pair states 
[see Fig.~\ref{fig:macrodimer_oscillations}(a) of the main text]. $H_\text{int}$ is chosen to have the 
same eigenenergies $V_\uparrow(r)$, $V_\downarrow(r)$ as the one obtained from exact diagonalization, 
and the same decomposition of its eigenstates over $\ket{+}$ and $\ket{+'}$ but neglecting residual 
overlaps with other pair states. The coupling term~$W(r)$ is then chosen to satisfy Eq.~(\ref{eq:V_up_down}). 
The resulting potentials and overlaps are represented in Fig.~\ref{fig:adiabaticity}(a).

To check that the adiabatic condition is satisfied, we perform a self-consistent analysis which 
considers classical nuclear motion and quantum electronic dynamics~\cite{Delos_1972,Billing_1983}. 
First, assuming that the adiabatic condition is satisfied, we simulate the time evolution of the average 
position~$r(t)$ under the 1D potential $V_\uparrow$ with initial condition $r_0 = 5a$~µm.
The resulting periodic trajectory is shown in Fig.~\ref{fig:adiabaticity}(b, bottom panel).
Then, we simulate the ideal time evolution of the internal state $\ket{\psi(t)}$ following the Schrödinger equation
\begin{align}
		i \hbar \frac{d\ket{\psi(t)}}{dt} = H_\text{int}\left[r(t)\right] \ket{\psi(t)},
\end{align}
with initial condition $\ket{\psi(t=0)} = \ket{\uparrow(r_0)} \simeq \ket{+}$. 
We show in Fig.~\ref{fig:adiabaticity}(b, top panel) the overlap~$P_\uparrow(t) = {|\langle\uparrow|\psi(t)\rangle|^2}$ 
on the instantaneous eigenstate $\ket{\uparrow}$ and check that it remains equal to 1 over the full dynamics, 
justifying \emph{a posteriori} the adiabatic approximation. 
Meanwhile, the overlap~$P_+(t) = {|\langle +| \psi(t) \rangle|^2}$ follows 
a periodic evolution [red curve in Fig.~\ref{fig:adiabaticity}(b)] that is synchronized with $r(t)$ 
and reaches a minimum of~$7$~\%.
	
The adiabatic condition can also be estimated from the Landau-Zener formula~\cite{Zener_1932}, 
if we make the approximation (valid only in order of magnitude in the vicinity of the avoided crossing) 
that $W(r)$ and $\frac{\partial}{\partial r} \left(V_+ - V_{+'}\right)$ do not depend on~$r$, 
and that the interatomic velocity~$dr / dt$ is constant. In this case, the probability of a 
diabatic transfer from $\ket{\uparrow}$ to $\ket{\downarrow}$ after going through 
the avoided crossing once reads~\cite{Zener_1932,Wittig_2005}
\begin{align}
	P_{\ket{\uparrow}\rightarrow\ket{\downarrow}} = e^{-2\pi \Gamma} \quad
	\text{with} \quad \Gamma = \dfrac{|W|^2}{\left|\dfrac{dr}{dt}\right| \cdot \left|\dfrac{\partial
	\left(V_+ - V_{+'}\right)}{\partial r} \right|}
\end{align}
With $W \simeq 2\pi \times 5$~MHz, $\frac{\partial}{\partial r} \left(V_+ - V_{+'}\right) \simeq 2\pi \times 21$~MHz$/$µm 
and $|dr / dt| \simeq 0.39$~m$/$s at the avoided crossing, we find $\Gamma \simeq 19$ 
leading to a completely negligible probability~$P_{\ket{\uparrow}\rightarrow\ket{\downarrow}}$.

\bibliography{spin_motion_coupling_biblio}

\end{document}